\documentclass[tightenlines,superscriptaddress,eqsecnum,floats,nofootinbib,showpacs]
{revtex4}
\usepackage{amssymb}
\usepackage{stmaryrd}
\usepackage{amsmath}
\usepackage{amsfonts}
\usepackage{mathrsfs}

\usepackage{amsmath,amssymb,amsfonts}
\usepackage{graphicx}

\def\be{\begin{equation}}
\def\ee{\end{equation}}
\def\ba{\begin{eqnarray}}
\def\ea{\end{eqnarray}}
\def\nn{\nonumber}

\begin{document}


\title{ Thermodynamics in new model of loop quantum cosmology}

\author{Xiangdong Zhang\footnote{scxdzhang@scut.edu.cn}}
\affiliation{Department of Physics, South China University of
Technology, Guangzhou 510641, China}

\begin{abstract}

The thermodynamic properties of loop quantum cosmology (LQC) without considering the Lorentz term were established in \cite{Zhu09}. In this paper, we extend this result to the recent proposed new model of LQC with the Lorentz term. We investigate the thermodynamics of LQC on the apparent horizon of the Friedmann-Lematre-Robertson-Walker universe. The result shows that the effective density and effective pressure in the modified Friedmann equation of LQC not only determines the evolution of the universe but can also serve as the thermodynamic quantities. Moreover, with the help of the Misner-Sharp energy, the first law of thermodynamics of the LQC is still valid as expected. This in turn endows precise physical meaning to the effective matter density $\rho_{eff}$ and the effective pressure $P_{eff}$.

\end{abstract}


\maketitle

\section{Introduction}

As a nonperturbative
and background-independent quantization scheme
of gravity, loop quantum gravity (LQG) has been extensively investigated for the past three decades\cite{Ro04,Th07,As04,Ma07}. One of the most important and successful applications
of LQG is to study its symmetry-reduced model of cosmological case, which is usually referred to as loop quantum cosmology (LQC). The most remarkable feature of LQC is that it replaces the classical
singularities with a quantum bounce\cite{LQC5}, both in an isotropic and spatial flat model as well as in a
less symmetric homogeneous model\cite{Boj,Ash-view,AS11,BCM,APS3,ACS}. Moreover, it has been shown that the non-perturbative modification feature of LQC leads to a generic phase of inflation
\cite{Bojowald02,Ashtekar12}.

On the other hand, as pointed out by Hawking \cite{Hawking75}, the
thermodynamic properties of spacetime are coming from the quantum
effects of spacetime. Therefore, it is very interesting
and important to investigate the thermodynamics properties by taking into
account quantum gravity effects. There are much progress has been made in recent years on the issues of the unified first law of the universe\cite{Cai05,Cai09,Hu15,Hu11,Hayward98}. Also in the recent years, there are indeed  many authors are devoted to studying on
thermodynamics implications of loop quantum gravity with a particularly focusing on the black hole thermodynamics
\cite{Rovelli96,Perez11}. While on the LQC side, by considering that
the universe is being non-stationary and evolving, the thermodynamics
is then quite different from the black hole systems. Note that the evolution
of the early universe is drastically changed by the quantum gravity effect, there exists a quantum phase that is
dictated by a difference equation\cite{ACS}. In this stage, the
universe may not be an equilibrium system due to the fast quantum
evolution. Then, as the size of the universe increases
and matter density keep on decreases, the universe enters an intermediate
semi-classical phase in which the evolution of the universe can be well described by
the effective equations of LQC, and it is
reasonable to approximately treat the universe as a thermodynamic equilibrium
system. With this thermodynamic equilibrium
system of LQC in hand, the thermodynamics of LQC is constructed in \cite{Zhu09}. Similar to the black hole, the thermodynamics
of LQC are also subjected to the quantum effects of the universe.

However, the previous construction of the
framework of LQC is involving a particular choice of the Hamiltonian constraint operator. More precisely,
the Hamiltonian constraint in full LQG consists of two terms which are the Euclidean
term and the Lorentzian term respectively. In the classical spatially flat cosmological models,
the Lorentzian term is simply proportional to the Euclidean term. Thus one
could simplify these two terms into a single term, and then quantize it by the methods inherited from the full theory of LQG to obtain the well-defined Hamiltonian constraint operator in
the cosmological models. This standard LQC coupled with massless scalar field leads to
the symmetric bounce of the Universe \cite{ACS,Boj}.

In contrast with the classical case, in the quantum situation, the Euclidean
term and Lorentzian term behave quite differently. In the full theory of LQG, these two terms were first regularized and quantized separately as operators in \cite{Thiemann06}. While in the cosmological case, this treatment
was first carried out in \cite{Ma09a}, where an alternative version of the Hamiltonian constraint operator was obtained in LQC. The notable feature of the
effective Hamiltonian of this alternative operator which was later confirmed by the semiclassical analysis of Thiemann's
Hamiltonian constraint operator in full LQG, is that it can lead to an asymmetric bounce scenario in LQC \cite{Pawlowski18,Pawlowski19,Wang18}. This new model of LQC now usually referred to as the Dapor-Liegener model of LQC, and the
remarkable result relates to the flat Friedmann-Lematre-Robertson-Walker(FLRW) cosmological spacetime with an asymptotic de Sitter
epoch. To inherit more features of this LQC model, and therefore get a better understanding of the thermodynamics of LQC, in this paper we will deal with the
Euclidean and the Lorentzian terms independently and consider the thermodynamics of LQC in this setting.

This paper is organized as follows: After an introduction, we will first give a brief review on LQC with Lorentz term and collecting the basic building blocks that we need in section \ref{Section2}.
Then in section \ref{Section3}, the detailed construction of thermodynamics of LQC with Lorentz term and the physical interpretation of the effective density and pressure will be given.
Our results will be summarized in the last section. Throughout the paper, we adopt the convention that the speed of light $c=1$.

\section{Brief review on Dapor-Liegener model of LQC}\label{Section2}

We consider the $k=0$, isotropic homogeneous $(3+1)$-dimensional LQC. According to\cite{Ma09,Pawlowski18}, the metric of the FLRW universe reads
\ba
ds^2=-dt^2+a(t)(dr^2+r^2d\Omega),\label{FRWmetric}
\ea where $a$ is the scale factor and $d\Omega$ denotes the standard $2$-dimensional sphere.

Note that in LQC, the conservation of energy-momentum tensor keeps the same form as in classical theory.  However, the Friedmann equation receives quantum correction as\cite{Pawlowski18,Pawlowski19}
\ba
H^2=\frac{\kappa}{3}\frac{\rho_{cL}}{(1+\gamma^2)}\left(1-\frac{\rho}{\rho_{cL}}\right)\left(1-\sqrt{1-\frac{\rho}{\rho_{cL}}}\right)\left(1+2\gamma^2+\sqrt{1-\frac{\rho}{\rho_{cL}}}\right)\label{effF}
\ea where $\rho$ is the matter density, $H=\frac{\dot{a}}{a}$ is the Hubble parameter, $\kappa=8\pi G$ and the new critical matter density $\rho_{cL}=\frac{\rho_c}{4(1+\gamma^2)}=\frac{3}{4\kappa (1+\gamma^2)\gamma^2\Delta}$ with $\Delta=4\sqrt{3}\pi\gamma G\hbar$ being the minimum area gap given by the full theory of LQG and $\gamma$ is the Barbero-Immirzi parameter \cite{As04,Ash-view}.
By introducing a function \ba
f(\rho)=\frac{1-\sqrt{1-\frac{\rho}{\rho_{cL}}}}{2(1+\gamma^2)},
\ea the above modified Friedmann equation can be rewritten in a compact form
\ba
H^2=\frac{1}{\gamma^2\Delta}f(\rho)\left(1-f(\rho)\right)\left(1-\frac{\rho}{\rho_{cL}}\right).
\ea
Compared with the classical Friedmann equation \ba
H^2&=&\frac{\kappa}{3}\rho.
\ea We can define the effective density\ba
\rho_{eff}=\frac{3}{\kappa\gamma^2\Delta}f(\rho)\left(1-f(\rho)\right)\left(1-\frac{\rho}{\rho_{cL}}\right).\label{rhoeff}
\ea In terms of this $\rho_{eff}$, the modified Friedmann equation can be written as
\ba
H^2&=&\frac{\kappa}{3}\rho_{eff}\label{Friedmanneff},
\ea The physical meaning of the $\rho_{eff}$ will become clear in the following. Moreover, it is of course clear from Eq. \eqref{rhoeff}, at the regime where the quantum corrections can be neglected (i.e. $\frac{\rho}{\rho_{cL}}\rightarrow 0$), $\rho_{eff}$ goes back to its classical value $\rho_{eff}=\rho$.

In addition, combining Eq.\eqref{effF} with the continuity equation
\ba
\dot{\rho}+3\frac{\dot{a}}{a}(P+\rho)=0,
\ea we obtain
the modified Raychaudhuri equation as
\ba
\frac{\ddot{a}}{a}=\dot{H}+H^2=\frac{3}{2\gamma^2\Delta}(\rho+P)f'(\rho)\left(2f(\rho)-1\right)\left(1-\frac{\rho}{\rho_{cL}}\right)
+\frac{1}{\gamma^2\Delta}f(\rho)\left(1-f(\rho)\right)\left(1+\frac{3P}{2\rho_{cL}}+\frac{\rho}{2\rho_{cL}}\right)\label{quantumRay}
\ea
Compared Eq. \eqref{quantumRay} with the standard Raychaudhuri equation \ba
\frac{\ddot{a}}{a}&=&-\frac{4\pi G}{3}(\rho+3P).
\ea We can define the effective pressure as \ba
P_{eff}=-\frac{3}{\kappa\gamma^2\Delta}\left[(\rho+P)f'(\rho)\left(2f(\rho)-1\right)\left(1-\frac{\rho}{\rho_{cL}}\right)
+f(\rho)\left(1-f(\rho)\right)\left(1+\frac{P}{\rho_{cL}}\right)\right]\label{Peff}
\ea where $f'(\rho)=\frac{df(\rho)}{d\rho}$. It is also easy to check that when $\frac{\rho}{\rho_{cL}}\rightarrow 0$, $P_{eff}$ goes back to its classical value $P_{eff}=P$.

With this effective density $\rho_{eff}$ and effective pressure $P_{eff}$ in hand, the modified Raychaudhuri
and conservation equations take the same form as their classical counterparts respectively \ba
\frac{\ddot{a}}{a}&=&-\frac{4\pi G}{3}(\rho_{eff}+3P_{eff}),\\
\dot{\rho}_{eff}&+&3\frac{\dot{a}}{a}(P_{eff}+\rho_{eff})=0\label{conservationeff}.
\ea At this stage, these similarities are nothing but mathematical
tricks to redefine the coupling of matter and gravity.
The complicated expression of Eqs.\eqref{rhoeff} and \eqref{Peff} made people hard believe these quantities can exist with a physically well-defined thermodynamic origin \cite{KD05}. To clarify these issues, we will explore their intrinsic
physical meaning in the thermodynamic sense in the following section and discuss
some further implications based on the above effective
framework of LQC with Lorentz term.

\section{Misner-Sharp energy and THERMODYNAMICS of LQC with the Lorentz term}\label{Section3}

For the cosmological situation, the evolution of
the universe is completely determined by the scale factor $a(t)$. Then the
metric \eqref{FRWmetric} can be rewritten as
\ba
ds^2=h_{ab}dx^adx^b+\tilde{r}^2d\Omega,\label{FRW2d}
\ea
where $h_{ab}=diag(-1,a^2)$ is the two dimensional
metric with $x^0=t, x^1=r$ and $\tilde{r}=a^2(t) r$
. And for the convenience of the following discussion, from now on, we will work in the convention such that $G=\hbar=1$.
For the FLRW universe, similar to the black hole case, it exists a horizon which is so-called the apparent horizon. Notably, the cosmological apparent horizon
is locally defined and dynamical, the boundary of the apparent horizon is a sphere that has vanishing expansion \cite{Cai05,Zhu09} as\ba
\theta=h^{ab}\partial_a\tilde{r}\partial_b\tilde{r}=0.\label{boundaryAH}
\ea The solution to Eq.\eqref{boundaryAH} gives rise to the radius of the apparent horizon \ba
R_A=\frac{1}{H}.\label{radiusAH}
\ea
Since we are considering thermodynamics, a quantity usually connected with the concept of temperature is the surface gravity $\kappa$. According to the definition\cite{Cai09,Hu11}, the surface gravity $\kappa$ on the apparent horizon of FLRW universe reads
\ba
\kappa=\frac{1}{2\sqrt{-h}}\partial_a\left(h^{ab}\partial_b\tilde{r}\right)=-\frac{1}{R_A}\left(1-\frac{\dot{R}_A}{2HR_A}\right),
\ea where the metric \eqref{FRW2d} has been used.

Note that we are discussing thermodynamics, then we need to specify the notion of energy that we used. However, in general relativity, the local energy density does not exist, the only well-defined version of energy is the quasi-local energy\cite{Hayward98}. There are many kinds of quasi-local energy, different versions of quasi-local energy have different merit. Among them, there exist a particular kind of quasi-local energy called as Misner-Sharp(MS) energy\cite{MSenergy} which is a purely geometric quantity and extensively used in the literature
about the thermodynamics of spacetime \cite{Hayward98,Cai07a,Cai07b}.
The physical meaning of MS energy and the comparison to the Arnowitt-Deser-Misner
mass and Bondi-Sachs energy can be found in \cite{Hayward96}. Now we introduce the Misner-Sharp energy $E_{MS}$, according to definition \cite{MSenergy,Cai09,Hayward96}, in the spherically symmetric
case and with unit $G=1$, this energy reads \ba
E_{MS}=\frac{r}{2}\left(1-h^{ab}\partial_ar\partial_br\right),
\ea
which is the total energy (not only the passive energy)
inside the sphere with radius $r$.

Now we consider the Misner-Sharp energy inside the apparent
horizon $r=R_A$ of the FLRW universe, a direct calculation shows
\ba
E_{MS}=\frac{R_A}{2}\label{RAEMS}.
\ea Note that in terms of the apparent horizon radius \eqref{radiusAH}, the modified Friedmann
equation \eqref{Friedmanneff} can be rewritten as \ba
\frac{1}{R^2_A}=\frac{8\pi }{3}\rho_{eff}.\label{RAFRWeff}
\ea Combine Eqs. \eqref{RAEMS} and \eqref{RAFRWeff} gives the total Misner-Sharp energy inside the apparent
horizon \ba
E_{MS}=\rho_{eff}V=\frac{4\pi R^3_A}{3}\rho_{eff}.\label{MSrhoV}
\ea The above equation endows $\rho_{eff}$ with a clear physical meaning. It shows that it is reasonable to say that $\rho_{eff}$ is indeed the
energy density rather than merely a mathematical symbol. Then
from the conservation law \eqref{conservationeff}, it is also reasonable
to call $P_{eff}$ as pressure. This in turn means that the gravitational
field also has a contribution to the energy density and pressure
in the thermodynamical sense. Now we will move on to figure out whether these physical interpretations are consistent with the
fundamental relation of thermodynamics or not.

To examine the fundamental relation of thermodynamics
in the framework of LQC with Lorentz term, we consider the apparent horizon
of the FLRW universe. In the early work, people usually assume that the apparent horizon also
have a temperature $T$ determined by its surface gravity \cite{Zhu09}, and later it was proved by Cai et.al.\cite{Cai09,Hu15} that there indeed has
a real thermal emission spectrum of Hawking temperature $T$ associated with the apparent horizon for a given surface gravity $\kappa$ as\ba
T=\frac{\kappa}{2\pi}.
\ea
By taking the derivative of the Eq. \eqref{MSrhoV}, and using
conservation equation \eqref{conservationeff}, we get
\ba
dE_{MS}=4\pi R^2_A\rho_{eff}\dot{R}_Adt-4\pi R^3_A H(\rho_{eff}+P_{eff})dt\label{dEMS}.
\ea
Meanwhile, by taking the derivative of the Friedmann Equation
(21) and using the conservation equation \eqref{conservationeff}, we
get the differential form of the Friedmann equation \ba
\frac{1}{R^3_A}dR_A=4\pi H(\rho_{eff}+P_{eff})dt.
\ea
Multiply both sides of the above equation
by a factor $R_A\left(1-\frac{\dot{R}_A}{2HR_A}\right)$ gives\ba
\frac{\kappa}{2\pi}d(\pi R^2_A)=-4\pi R^3_A H(\rho_{eff}+P_{eff})\left(1-\frac{\dot{R}_A}{2HR_A}\right)dt.
\ea With the help of this equation, Eq. \eqref{dEMS} could be written as
\ba
dE_{MS}&=&\frac{\kappa}{2\pi}d(\pi R^2_A)+\frac{(\rho_{eff}-P_{eff})}{2}dV\nn\\
&=&Td(\pi R^2_A)+\frac{(\rho_{eff}-P_{eff})}{2}dV\label{firstlaw}.
\ea Compare with the standard form of first law of thermodynamics\cite{Zhu09,Cai09} \ba
dE&=&TdS+WdV,
\ea we can easily identify the entropy $S$ of apparent horizon as \ba
S=\frac{A}{4}=\pi R^2_A,
\ea which is nothing but Bekenstein-Hawking entropy of the apparent horizon as expected. Moreover, the work density reads\ba
W=\frac{(\rho_{eff}-P_{eff})}{2}\label{workdensityeff}.
\ea The emergent of this work density is because that if we suppose the matter source in the FLRW universe is a perfect fluid with stress-energy tensor as \ba
T_{\mu\nu}=(\rho_{eff}+P_{eff})u_\mu u_\nu+P_{eff}g_{\mu\nu}.
\ea Following \cite{Hayward99}, we define the work density as \ba
W=-\frac12T^{\mu\nu}h_{\mu\nu},
\ea which could be regarded as the work done by the change of the apparent horizon. Moreover, it is worth to note that in the classical case, the work density reads \cite{Hayward98}
\ba
W=\frac{(\rho-P)}{2},
\ea which is nothing but the classical limit of our Eq.\eqref{workdensityeff}. Now, we can conclude that
the physical interpretation of $\rho_{eff}$ and $P_{eff}$ in LQC as
thermodynamical quantities is consistent with
the fundamental relation of thermodynamics. In other words,
the fundamental relation of thermodynamics is valid in
the new version of LQC with respect to thermodynamical quantities $\rho_{eff}$  and $P_{eff}$.

\section{Conclusion}\label{Section5}

In summarize, we study the thermodynamic
properties of the universe in LQC with the Lorentz term and
found that the fundamental relation of thermodynamics
is still valid.
In order to investigate
the thermodynamic properties at the apparent horizon
of the FLRW universe. We develop a procedure based on the effective theory of LQC with the Lorentz term and with the
homogeneous and isotropic cosmological setting. By introducing
the MS energy, and using the thermodynamic quantities associated with the cosmological apparent horizon, the first law of thermodynamics of the LQC is expected to be valid.
The resulted work density $W$ also has the correct classical limit.

It is interesting to found that although the $\rho_{eff}$ and $P_{eff}$ have complicated expressions, it turns out that the
effective matter density $\rho_{eff}$ and the effective pressure $P_{eff}$
possess real physical meaning rather than merely being complicated mathematical symbols. More precisely, the $\rho_{eff}$
being the density of MS energy inside the spherically symmetric apparent horizon. In the large volume or the spatial
curvature is negligible, the $\rho_{eff}$ and $P_{eff}$ approach to its classical value of matter density $\rho$ and pressure $P$.

There are of course a few issues that deserve further investigating based on our thermodynamic of LQC. First, it is interesting to generalize the result presented in this paper to the other theories of gravity, which would be helpful to get a better understanding of the thermodynamic properties of a more general version of quantum gravity. Second, since we endow $\rho_{eff}$ and $P_{eff}$ with precise physical meaning, it is therefore very interesting to found other applications of such quantities. For example, is it possible to construct the wormhole solution with such $\rho_{eff}$ in effective LQC? We would like to leave these interesting topics for the future.

\begin{acknowledgements}
This work is supported by NSFC with No.11775082

\end{acknowledgements}


\begin{thebibliography}{99}


\bibitem{Zhu09}L. Li and J. Zhu, {\it Thermodynamics in Loop Quantum Cosmology}, Adv.High Energy Phys. 2009:905705, (2009).

\bibitem{Ro04} C. Rovelli, Quantum Gravity, (Cambridge University Press, 2004).

\bibitem{Th07} T. Thiemann, Modern Canonical Quantum General Relativity, (Cambridge University
Press, 2007).


\bibitem{As04}A. Ashtekar and J. Lewandowski, {\it Background independent quantum gravity: A
  status report}, Class. Quant. Grav. {\bf21}, R53 (2004).

\bibitem{Ma07} M. Han, Y. Ma and W. Huang, {\it Fundamental structure of loop quantum gravity}, Int. J. Mod. Phys. D {\bf16}, 1397 (2007).

\bibitem[Ashtekar et~al.(2003)Ashtekar, Bojowald, Lewandowski, et~al.]{LQC5}
A.~Ashtekar, M.~Bojowald, J.~Lewandowski.
\newblock Mathematical structure of loop quantum cosmology.
\newblock \emph{Advances in Theoretical and Mathematical Physics}, 7\penalty0
  (2):\penalty0 233--268, 2003.

\bibitem[Bojowald(2008)]{Boj}
M.~Bojowald.
\newblock Loop quantum cosmology.
\newblock \emph{Living Reviews in Relativity}, 11\penalty0 (1):\penalty0 4,
  2008.

\bibitem[Ashtekar(2009)]{Ash-view}
A.~Ashtekar.
\newblock Loop quantum cosmology: an overview.
\newblock \emph{General Relativity and Gravitation}, 41\penalty0 (4):\penalty0
  707--741, 2009.

\bibitem[Ashtekar and Singh(2011)]{AS11}
A.~Ashtekar and P.~Singh.
\newblock Loop quantum cosmology: a status report.
\newblock \emph{Classical and Quantum Gravity}, 28\penalty0 (21):\penalty0
  213001, 2011.

\bibitem[Banerjee et~al.(2012)Banerjee, Calcagni, Martin-Benito, et~al.]{BCM}
K.~Banerjee, G.~Calcagni, M.~Martin-Benito, et~al.
\newblock Introduction to loop quantum cosmology.
\newblock \emph{SIGMA. Symmetry, Integrability and Geometry: Methods and
  Applications}, 8:\penalty0 016, 2012.

\bibitem[Ashtekar et~al.(2006)Ashtekar, Pawlowski, and Singh]{APS3}
A.~Ashtekar, T.~Pawlowski, and P.~Singh.
\newblock Quantum nature of the big bang: improved dynamics.
\newblock \emph{Physical Review D}, 74\penalty0 (8):\penalty0 084003, 2006.

\bibitem[Ashtekar et~al.(2008)Ashtekar, Corichi, and Singh]{ACS}
A.~Ashtekar, A.~Corichi, and P.~Singh.
\newblock Robustness of key features of loop quantum cosmology.
\newblock \emph{Physical Review D}, 77\penalty0 (2):\penalty0 024046, 2008.

\bibitem{Bojowald02}M. Bojowald, {\it Inflation from Quantum Geometry}, Phys. Rev. Lett. 89, 261301 (2002).

\bibitem{Ashtekar12}I. Agullo, A. Ashtekar, W. Nelson, {\it A Quantum Gravity Extension of the Inflationary Scenario}, Phys. Rev. Lett. 109, 251301 (2012).

\bibitem{Hawking75}S. Hawking, Commun. Math. Phys. 43, 199 (1975).

\bibitem{Cai05}R. G. Cai and S. P. Kim, {\it First Law of Thermodynamics and Friedmann Equations of Friedmann-Robertson-Walker Universe}, J. High Energy Phys. 02 (2005).
050.

\bibitem{Cai09}R. Cai, L. Cao, Y. Hu, {\it Hawking Radiation of Apparent Horizon in a FRW Universe}, Class.Quant.Grav. 26:155018, (2009).

\bibitem{Hu15}Y. Hu, H. Zhang, {\it Misner-Sharp Mass and the Unified First Law in Massive Gravity}, Phys.Rev. D 92,  024006 (2015).


\bibitem{Hu11}Y. Hu, {\it Hawking radiation from the cosmological horizon in a FRW universe}, Phys.Lett.B701: 269-274, (2011).

\bibitem{Hayward98}S. A. Hayward, {\it Unified first law of black-hole dynamics and relativistic thermodynamics}, Class. Quantum Grav. 15, 3147 (1998).








\bibitem{Rovelli96}C. Rovelli, {\it Black Hole Entropy from Loop Quantum Gravity}, Phys. Rev. Lett. 77, 3288 (1996).


\bibitem{Perez11}A. Ghosh and A. Perez, {\it Black hole entropy and isolated horizons thermodynamics}, Phys. Rev. Lett. 107, 241301 (2011).



\bibitem{Thiemann06}T. Thiemann, {\it The phoenix project: master constraint programme for loop quantum gravity}, Classical and
Quantum Gravity, 23(7):2211, 2006.

\bibitem{Ma09a}J. Yang, Y. Ding, and Y. Ma, {\it Alternative quantization of the hamiltonian in loop quantum cosmology}, Physics
Letters B, 682(1):1-7, 2009.

\bibitem{Pawlowski18}M. Assanioussi, A. Dapor, K. Liegener, and T. Pawlowski, {\it Emergent de sitter epoch of the quantum cosmos
from loop quantum cosmology}, Physical Review Letters, 121(8):081303, 2018.


\bibitem{Pawlowski19}M. Assanioussi, A. Dapor, K. Liegener, and T. Pawlowski, {\it Emergent de Sitter epoch of the Loop Quantum Cosmos: a detailed analysis},  Phys. Rev. D 100, 084003 (2019).


\bibitem{Wang18}B.-F. Li, P. Singh, and A. Wang, {\it Towards cosmological dynamics from loop quantum gravity}, Physical Review
D, 97(8):084029, 2018.

\bibitem{KD05}K. Banerjee and G. Date, {\it Discreteness Corrections to the Effective Hamiltonian of Isotropic Loop Quantum Cosmology}, Class. Quantum Grav. 22, 2017.
(2005).



\bibitem{MSenergy}C. Misner, and D. Sharp, {\it Relativistic Equations for Adiabatic, Spherically Symmetric Gravitational Collapse},d Phys. Rev. 136, B571 (1964).

\bibitem{Cai07a} M. Akbar and R. G. Cai, {\it Thermodynamic behavior of the Friedmann equation at the apparent horizon of the FRW universe}, Phys. Rev. D 75, 084003 (2007).

\bibitem{Cai07b}R. G. Cai and L. M. Cao, {\it Unified first law and the thermodynamics of the apparent horizon in the FRW universe}, Phys. Rev. D 75, 064008
(2007).

\bibitem{Hayward96}S. A. Hayward, {\it Gravitational energy in spherical symmetry}, Phys. Rev. D 53, 1938 (1996)

\bibitem{Ma09}J. Yang, Y. Ma, {\it Quasi-Local Energy in Loop Quantum Gravity}, Phys. Rev. D 80: 084027,(2009).

\bibitem{Hayward99}S. A. Hayward, S. Mukohyana, and M. C. Ashworth. {\it Dynamic black-hole entropy}, Phys. Lett. A 256, 347 (1999)



\end{thebibliography}
\end{document}